\documentclass[11pt]{article}
\usepackage{amsmath}
\usepackage{graphicx}
\usepackage{amsfonts}
\usepackage{amssymb}
\newcommand{\ben}{\begin{enumerate}}
\newcommand{\een}{\end{enumerate}}
\newcommand{\be}{\begin{equation}}
\newcommand{\ee}{\end{equation}}
\newcommand{\bse}{\begin{subequation}}
\newcommand{\ese}{\end{subequation}}
\newcommand{\bea}{\begin{eqnarray}}
\newcommand{\eea}{\end{eqnarray}}
\newcommand{\bc}{\begin{center}}
\newcommand{\ec}{\end{center}}

\begin{document}
\author{\textbf{Maia Angelova}$^{\dagger}$ \textbf{\ and A.
Frank}$^{\ddagger}$\\$^{\dagger}$\textit{School of Computing and
Mathematics}\\\textit{University of Northumbria, Newcastle upon
Tyne, UK GB-NE1 8ST}\\$^{{}}\ddagger$\textit{Instituto de Ciencias
Nucleares and Centro de Ciencias F\'{i}sicas,}\\\textit{UNAM, A.P.
70-543, Mexico, D.F., 04510 Mexico.}\\}
\title{Algebraic Approach to Molecular Thermodynamics}
\maketitle
\begin{abstract}
An algebraic model based on Lie-algebraic and discrete symmetry techniques is
applied to the analysis of thermodynamic vibrational properties of molecules.
The local anharmonic effects are described by a Morse-like potential and the
corresponding anharmonic bosons are associated with the $SU(2)$ algebra. A
vibrational high-temperature partition function and the related thermodynamic
functions are derived and studied in terms of the parameters of the model. The
idea of a critical temperature is introduced in relation with the specific
heat. A physical interpretation of a quantum deformation associated with the
model is given.

\end{abstract}


\section{Introduction}

The algebraic approach has been used successfully in molecular physics and has
led to new insights into the nature of complex many body systems
\cite{iach:interacting,frank:algebraic,iach:algebraic}. In the framework of
the algebraic method, the Hamiltonian of a given system is written as an
algebraic operator using the generators of the appropriate Lie algebra. All
other operations in the model are algebraic operators, unlike the differential
operators in standard wave mechanics. The technical advantage of the algebraic
approach is the comparative ease of algebraic operations. Equally important,
however, is the conclusion derived from comparison with experiment, that there
are generic forms of symmetry-adapted algebraic Hamiltonians and that entire
classes of molecules can be described by generic Hamiltonians where the
parameters vary in a systematic fashion for different molecules. In its
initial stage of development \cite{iach:chem}- \cite{lemus:chem}, the
algebraic approach has sought to show why and how it provides a framework for
the understanding of large-amplitude anharmonic motion. The anharmonicities
are introduced by means of dynamical groups that correspond to anharmonic
potentials and which constrain the total number of levels to a finite value.
Later on, the $SU(2)$ models \cite{frank:annals}- \cite{veinte} combined Lie
algebraic techniques, describing the interatomic interactions, with discrete
symmetry techniques associated with the local symmetry of the molecules.
Recently, a clear-cut connection could be established between the
Morse-$SU(2)$ approach and the traditional potential energy surface methods
\cite{frank:chem,car:2000,lemus:chem}.

The algebraic model \cite{iach:interacting,frank:algebraic} exploits the
isomorphism between the $SU(2)$ algebra and the one-dimensional Morse
oscillator,
\begin{equation}
\mathcal{H}=-\frac{\hbar^{2}}{2\mu}\frac{d^{2}}{dx^{2}}+D(1-e^{-\frac{x}{d}
})^{2},
\end{equation}
where $D$ is the depth of the potential well, $d$ is its width, $x$ is the
displacement from the equilibrium and $\mu$ is the mass of the oscillator. The
one-dimensional Morse Hamiltonian can be written in terms of the generators of
$SU(2)$,
\begin{equation}
\mathcal{H}_{M}=\frac{A}{4}\left(  \hat{N}^{2}-4\hat{J}_{Z}^{2}\right)
=\frac{A}{2}(\hat{J}_{+}\hat{J}_{-}+\hat{J}_{-}\hat{J}_{+}-\hat{N}),
\end{equation}
where $A$ is a constant dependent on the parameters of the Morse potential.
The eigenstates, $|\!\lfloor{N}\rfloor,v\rangle$, correspond to the
$\ U(2)\supset SU(2)$ symmetry-adapted basis,$\;$where $N\;$is the total
number of bosons fixed by the potential shape, and $v$ is the number of quanta
in the oscillator, $v=1,2,\ldots,\left\lfloor \frac{N}{2}\right\rfloor $. The
value of $N$ is dependent on the depth $D$ and the width $d$ of the Morse
potential well \cite{iach:interacting,frank:algebraic, frank:annals},
\begin{equation}
N+1=\left(  \frac{8\mu Dd^{2}}{\hbar^{2}}\right)  ^{\frac{1}{2}}.\label{ND}%
\end{equation}
The parameters $A$ and $N$ are related to the usual harmonic and anharmonic
constants $\omega_{e}$ and $x_{e}\omega_{e}$ used in spectroscopy
\cite{{iach:chem},{levine:chem},{herzberg:spectra}},
\begin{align}
\label{wx}\omega_{e}  &  =A(N+1)=\hbar\left(  \frac{2D}{\mu d^{2}}\right)
^{\frac{1}{2}},\\
x_{e}\omega_{e}  &  =A=\frac{\hbar^{2}}{2d^{2}D}.\nonumber
\end{align}
The anharmonic effects can be described by anharmonic boson operators
\cite{frank:algebraic},
\begin{equation}
\hat{b}=\frac{\hat{J}_{+}}{\sqrt{N}},\;\;\;\hat{b}^{\dagger}=\frac{\hat{J}
_{-}}{\sqrt{N}}\;,\;\;\;\hat{v}=\frac{\hat{N}}{2}-\hat{J}_{z}%
\end{equation}
where $\hat{v}\;\;$is the Morse phonon operator with an eigenvalue $v$. The
operators satisfy the commutation relations,
\begin{equation}
\left[  \hat{b},\hat{v}\right]  =\hat{b},\;\;\;\;\;\;\;\left[  \hat
{b}^{^{\dagger}},\hat{v}\right]  =-\hat{b}^{^{\dagger}},\;\;\;\;\left[
\hat{b},\hat{b}^{^{\dagger}}\right]  =1-\frac{2\hat{v}}{N}\label{anh}%
\end{equation}
The harmonic limit is obtained when $\;\;N\rightarrow\infty$, \ in which case
\ \ $\left[  \hat{b},\hat{b}^{^{\dagger}}\right]  \!\!\rightarrow\!\!1$
\ giving the usual boson commutation relations.

The Morse Hamiltonian can be written in terms of \ the operators $\hat{b}$ and
$\hat{b}^{^{\dagger}}$,
\begin{equation}
H_{M}\sim\frac{1}{2}\left(  \hat{b}\hat{b}^{^{\dagger}}+\hat{b}^{^{\dagger}%
}\hat{b}\right)
\end{equation}
which corresponds to vibrational energies
\begin{equation}
\varepsilon_{v}=\hbar\omega_{0}\left(  v+\frac{1}{2}-\frac{v^{2}}{N}\right)
,\;v=1,2,\ldots,\left[  \frac{N}{2}\right] \label{Ev}%
\end{equation}
where $\omega_{0}$ is the harmonic oscillator frequency. Thus, the spectrum of
the Morse potential leads to a deformation of the harmonic oscillator algebra.
A more detailed relationship between the Morse coordinates and momenta and the
$SU(2)$ generators can be derived through a comparison of their matrix
elements \cite{frank:chem} and through the derivation of raising and lowering
operators for the Morse potential \cite{car:2000}. Note that for an infinite
potential depth, $N\!\rightarrow\!\infty$, the Morse potential cannot be
distinguished from the harmonic potential. To sumarize, the algebraic
anharmonic model has been developed to analyze molecular vibrational spectra
\ \cite{frank:algebraic}-\cite{veinte}. It provides a systematic procedure for
studying vibrational excitations in a simple form by describing the stretching
and bending modes in a unified scheme based on $SU(2)$ algebras which
incorporate the anharmonicity at the local level.

The deformation of the harmonic oscillator algebra, associated with the Morse
potential, has also been derived using a quantum analogue for the anharmonic
oscillator \cite{angelova:JPA}. We have described the anharmonic vibrations as
anharmonic $q$-bosons using a first-order expansion of the quantum
deformation. We have thus proposed a physical interpretation of quantum
deformation in the framework of the algebraic model.

The aim of this paper is to apply the algebraic approach to the vibrational
high-temperature thermodynamics of diatomic molecules and to obtain the basic
thermodynamic functions in terms of the parameters of the algebraic model.
This paper can be considered as a first step in the direction of incorporating
anharmonicity and the finite number of bound vibrational states into the
thermodynamic description of molecular systems. The rest of the paper is
organized as follows. In Section 2, we derive a Morse-like vibrational
partition function for high temperature and study its properties. In Section
3, the vibrational partition function is used to derive the basic
thermodynamic functions, such as the mean vibrational energy, specific heat
and free energy. The idea of critical temperature is introduced in relation
with the specific heat. In Section 4, the mean number of anharmonic bosons is
obtained. The concept of maximal temperature of the anharmonic vibrations is
discussed. The $q$-bosonic deformation of first order is considered. It is
shown that this quantum deformation is related to the shape of the anharmonic
potential well and the fixed number of anharmonic bosons. The results are
applied to the diatomic molecule H$^{1}$Cl$^{35}$. These results must be
combined with the translational and rotational thermodynamic functions in
order to compare with experiment, as discussed in reference
\cite{herzberg:spectra}.

\section{Vibrational Partition Function}

The vibrational partition function of a diatomic anharmonic molecule is
\begin{equation}
\label{def}Z_{N}=\sum_{v=0}^{[N/2]}e^{-\beta\varepsilon_{v}}%
\end{equation}
where $\beta=1/k_{B}T$, the vibrational energies $\varepsilon_{v}$ are given
by equation (\ref{Ev}) and $N$ is the fixed total number of anharmonic bosons.
Introducing new parameters,$\;\alpha=\frac{\beta\hbar\omega_{0}}{2}%
,\;N_{0}=\left\lfloor \frac{N}{2}\right\rfloor $ and $\;l=N_{0}-v$, the exact
value of vibrational partition function can be written as,
\begin{equation}
Z_{N}=e^{-\alpha}\sum_{l=0}^{N_{0}}e^{-\frac{\alpha}{N_{0}}\left(  N_{0}%
^{2}-l^{2}\right)  }.\label{exactpf}%
\end{equation}
At high temperatures $T$, for $N_{0}$ large and $\alpha<1$, the sum can be
replaced by the integral,
\begin{equation}
Z_{N}=\sqrt{\frac{N_{0}}{\alpha}}e^{-\alpha\left(  N_{0}+1\right)  }%
\int\limits_{0}^{\sqrt{\alpha N_{0}}}e^{s^{2}}ds
\end{equation}
where $s=\sqrt{\frac{\alpha}{N_{0}}}l$. This integral can be evaluated exactly
in terms of the error function, ${\hbox{\rm
erf} \,}i\!\left(  \sqrt{{\alpha
}N_{0}}\right)  $ (as defined in \cite{abram:handbook}),
\begin{equation}
Z_{N}=\frac{1}{2}\sqrt{\frac{N_{0}{\pi}}{\alpha}}e^{-\alpha\left(
N_{0}+1\right)  }{\hbox{\rm erf} \,}i\!\left(  \sqrt{\alpha N_{0}}\right)
.\label{pf}%
\end{equation}

\begin{figure}[t]
\input{epsf}  \centerline{\epsffile{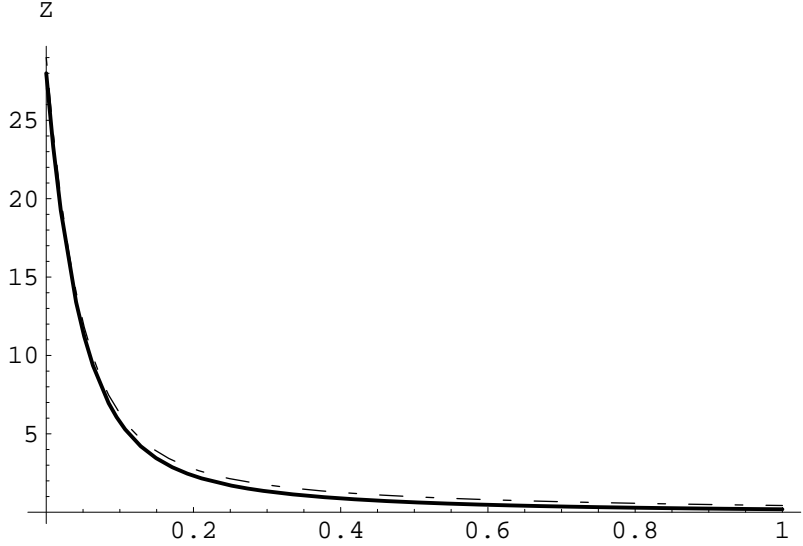}} \caption{Vibrational partition
function $Z_{56}$ as a function of $\alpha$. The double-dashed line represents
the exact representation.}%
\label{fig:vibra1}%
\end{figure}

Equation (\ref{pf}) represents the high-temperature value of the vibrational
partition function in the Morse-like spectrum
\cite{angelova:trjphys,angelova:CMP,angelova:Kra}. The partition function is
expressed in terms of the parameters of the algebraic model $N_{0}$ and
$\alpha$. The dependence on the temperature is given by $\alpha$,
\begin{equation}
\alpha=\frac{\hbar\omega_{0}}{2k_{B}T}=\frac{\Theta}{2T}.
\end{equation}
where $\Theta=\hbar\omega_{0}/k_{B}$ is the usual characteristic vibrational
temperature of the molecule. The contributions of the anharmonic vibrations
are essential in the high-temperature region for $T\geq\Theta$, where
$T=\Theta$ corresponds to $\alpha=0.5$.

When $N_{0}\rightarrow\infty$, the harmonic limit of the model is obtained,
\begin{equation}
Z_{\infty}\sim\frac{N_{0}e^{-\alpha}}{2\alpha N_{0}-1}\sim\frac{e^{-\alpha}%
}{2\alpha}==\frac{T}{\Theta}e^{-\frac{\Theta}{2T}}\label{Zhl}%
\end{equation}
which coincides with the harmonic vibrational partition function of a diatomic
molecule at high temperatures. The expression for the partition function
(\ref{pf}) can be generalised to polyatomic molecules, by combining the
present results with the use of a local-mode model where each interatomic
potential is of the Morse form \cite{lemus:chem}.

\begin{figure}[t]
\input{epsf}  \centerline{\epsffile{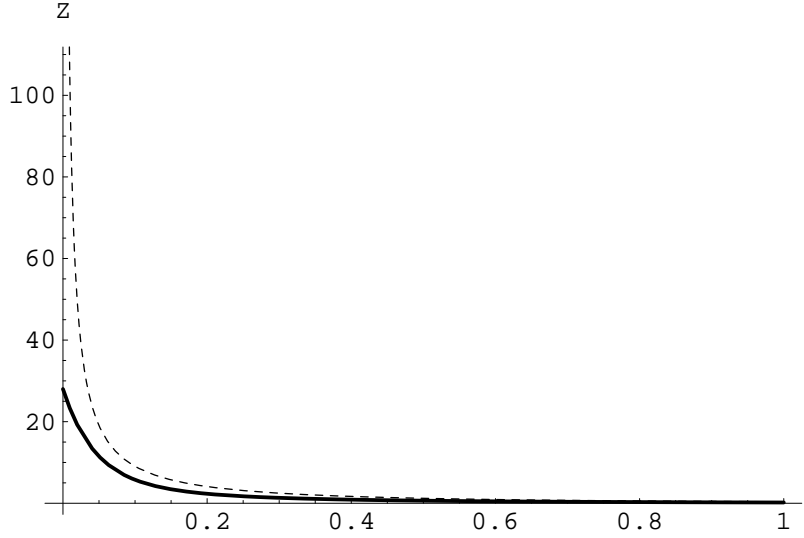}} \caption{ Vibrational partition
function $Z_{56}$ (solid line) and the harmonic limit $Z_{\infty}$ (dashed
line) as functions of $\alpha$.}%
\label{fig:vibra2}%
\end{figure}

The diatomic molecule H$^{1}$Cl$^{35}$ is considered as an example. Using the
values of $x_{e}$ and $\omega_{e}$ \cite{herzberg:spectra} for the zero lines
of this molecule and equations (\ref{ND}) and (\ref{wx}) we obtain the total
number of anharmonic bosons, fixed by the shape of the Morse potential,
$N=56$, and the total number of quanta in the oscillator, $N_{0}=28$. The
depth of the Morse potential is $D=5.32$eV, and the width is $d=0.57\times
10^{-10}$m. The characteristic vibrational temperature of the molecule is
$\Theta=4300$K \cite{BS:statmech}.

Substituting the value of $N_{0}=28$ in equation (\ref{pf}) we can calculate
the partition function, $Z_{56}$, for the molecule H$^{1}$Cl$^{35}$ as a
function of the parameter $\alpha$. The graph on Figure \ref{fig:vibra1}
represents the partition function $Z_{56}$ given by equation (\ref{pf}) for
the values of the parameter $\alpha$ between 0 and 1 (solid line). The exact
partition function from equation (\ref{exactpf}) is given for comparison
(double-dashed line). It is clear that in the region $0<\alpha\leq1$, the
integral approximation is in a very good agreement with the exact
representation and does not change the value and appearance of the partition
function. The comparison between the function $Z_{56}$ (solid line) and the
harmonic limit $Z_{\infty}$ (dashed line) is given on Figure \ref{fig:vibra2}.
The finiteness of $Z_{56}$ in the high $T$ limit is linked, of course, with
the finite number of states in the Morse potential. A more realistic
description in the high $T$ region requires the introduction of the continuum
states of the Morse potential \cite{ADF:prep}.

Having the partition function $Z_{N}$ in terms of the parameters of the
algebraic model, we are now in position to derive the corresponding
thermodynamic functions. An algebraic approach has been used in
\cite{kusnetzov:chem} to study the thermodynamic properties of molecules.
However, the partition function in \cite{kusnetzov:chem} uses an approximation
of the classical density of states, while we have derived an explicit function
in terms of the parameters of the algebraic model.

\section{Thermodynamic Vibrational Functions}

\subsection{Mean Vibrational Energy}

The mean vibrational energy is given by
\begin{equation}
U_{N}=-\frac{\partial}{\partial\beta}\hbox{\rm
ln} Z_{N}=-\frac{\hbar
\omega_{0}}{2Z_{N}}\frac{\partial Z_{N\;}}{\partial\alpha}.
\end{equation}
Taking into account that
\begin{equation}
\frac{\partial Z_{N\;}}{\partial\alpha}=-\frac{Z_{N}}{2\alpha}-\left(
N_{0}+1\right)  Z_{N}+\frac{N_{0}e^{-\alpha}}{2\alpha},\label{dZ}%
\end{equation}
we obtain the following expression for the mean vibrational energy in terms of
the partition function $Z_{N}$,
\begin{equation}
U_{N}=\frac{\hbar\omega_{0}}{2}\left(  1+N_{0}+\frac{1}{2\alpha}%
-\frac{N_{0}e^{-\alpha}}{2\alpha Z_{N}}\right)  .\label{UN}%
\end{equation}
Substituting $Z_{N}$ with equation (\ref{pf}) gives the following expression
for the mean energy, $U_{N}$, in terms of the parameter ${\alpha}$,
\begin{equation}
U_{N}=\frac{\hbar\omega_{0}}{2}\left(  1+N_{0}+\frac{1}{2\alpha}%
-\sqrt{\frac{N_{0}}{\alpha\pi}}\frac{e^{\alpha N_{0}}}{\hbox{\rm
erf}
\,i\left(  \sqrt{\alpha N_{0}}\right)  }\right)  .\label{U}%
\end{equation}
The harmonic limit is obtained from equation (\ref{UN}), when $N_{0}$
$\rightarrow\infty$ and $Z_{N}$ is given by (\ref{Zhl}),
\begin{equation}
U_{\infty}\sim\frac{\hbar\omega_{0}}{2}(1+\frac{1}{\alpha})=\frac{\hbar
\omega_{0}}{2}+k_{B}T.
\end{equation}
This is the classical mean energy of a diatomic molecule at high temperatures.
\begin{figure}[t]
\input{epsf}  \centerline{\epsffile{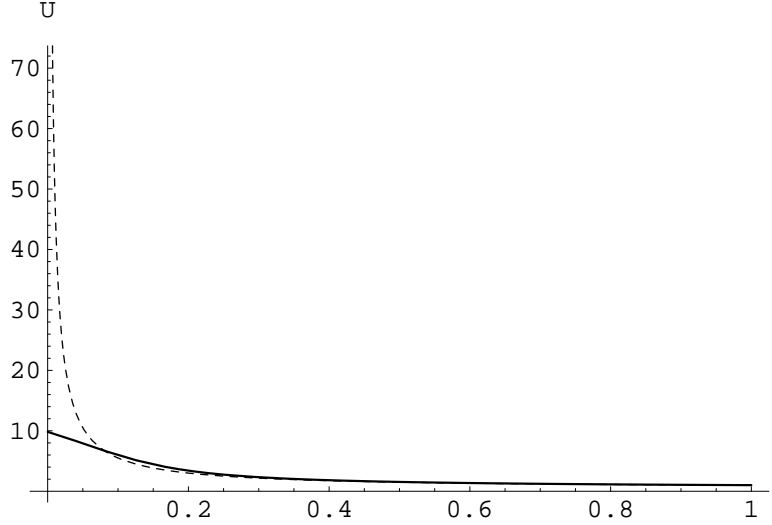}} \caption{Mean vibrational energy
$U_{56}/{\hbar\omega_{0}}$ as a function of $\alpha$. The dashed line
represents the harmonic limit $U_{\infty}/{\hbar\omega_{0}}$.}%
\label{fig:energy}%
\end{figure}

The graph on Figure \ref{fig:energy} represents the mean vibrational energy,
$U_{56}/{\hbar\omega_{0}}$, of the molecule H$^{1}$Cl$^{35}$ for values of
$\alpha$, $0<\alpha\leq1$. The high-temperature region corresponds to
$\alpha\leq0.5$. For comparison the graph of the harmonic limit $U_{\infty}$
(dashed line) for $N_{0}\rightarrow\infty$ is also given. As already mentioned
in the previous section, the finiteness of $U_{N}$ in the high $T$ limit is a
result of the finite number of states in the Morse potential.

\subsection{Specific Heat}

The vibrational part of the specific heat is,
\begin{equation}
C_{N}=\frac{\partial U_{N}}{\partial T}=-\frac{\hbar\omega_{0}}{2k_{B}T^{2}%
}\frac{\partial U_{N}}{\partial\alpha}.
\end{equation}
Substituting $U_{N}$ with equation (\ref{UN}) and using equation (\ref{dZ}),
we obtain
\begin{equation}
C_{N}=\frac{k_{B}}{2}+\frac{k_{B}N_{0}e^{-\alpha}}{2Z_{N}}\left(  \alpha
N_{0}-\frac{1}{2}-\frac{N_{0}e^{-\alpha}}{2Z_{N}}\right) \label{CN}%
\end{equation}
This equation represents the vibrational specific heat in the algebraic model
in terms of the partition function $Z_{N}$. Substituting $Z_{N}$ in equation
(\ref{CN})with the expression (\ref{pf}), we obtain the dependence of the
specific heat $C_{N}$ on the parameter ${\alpha}$,
\begin{equation}
C_{N}=\frac{k_{B}}{2}+k_{B}\sqrt{\frac{\alpha N_{0}}{\pi}}\frac{e^{\alpha
N_{0}}}{{\hbox{\rm erf} \,}i\left(  \sqrt{\alpha N_{0}}\right)  }\left(
\alpha N_{0}-\frac{1}{2}-\sqrt{\frac{\alpha N_{0}}{\pi}}\frac{e^{\alpha N_{0}%
}}{\hbox{\rm erf} \,i\left(  \sqrt{\alpha N_{0}}\right)  }\right) \label{C}%
\end{equation}
It is clear from the relation (\ref{C}) that all anharmonic contributions to
the vibrational part of the specific heat depend on the parameter $\alpha$ and
hence on the temperature $T$. When $N_{0}~$ $\rightarrow\infty$, the harmonic
limit of the model gives the vibrational specific heat of a diatomic molecule
at high temperatures,
\begin{equation}
C_{\infty}\sim k_{B}.\label{CHL}%
\end{equation}

\begin{figure}[t]
\input{epsf}  \centerline{\epsffile{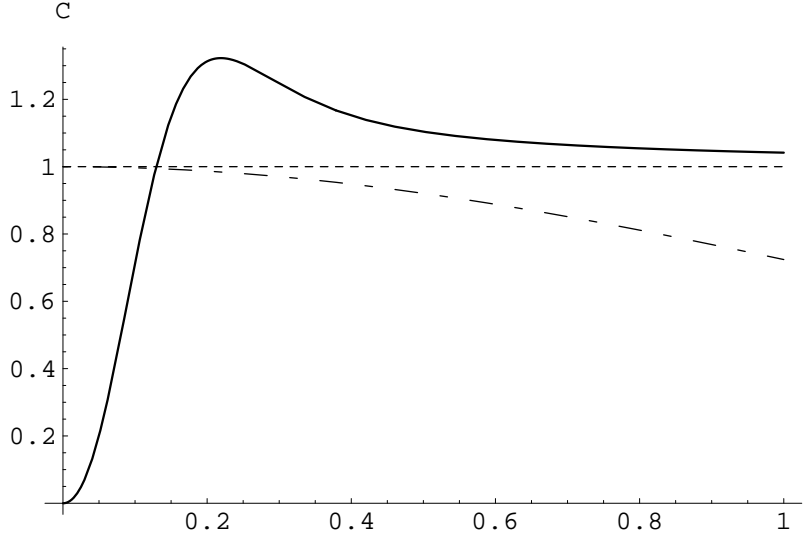}} \caption{Vibrational specific
heat $C_{56}/k_{B}$ (solid line) as a function of $\alpha$. For comparison,
$C_{harm}/k_{B}$ (double-dashed line) and $C_{\infty}/k_{B}$ (dashed line) are
also given.}%
\label{fig:heat}%
\end{figure}Figure \ref{fig:heat} represents the dependence of the vibrational
specific heat, $C_{56}/k_{B}$, on the parameter ${\alpha}$, $0<\alpha<1$, for
the molecule H$^{1}$Cl$^{35}$ (solid line). The graph of the harmonic
vibrational specific heat of a diatomic molecule (see {.g.}
\cite{herzberg:spectra,BS:statmech}), $C_{harm}/k_{B}$, is also given
(double-dashed line) as well as the harmonic limit, $C_{\infty}/k_{B}$ (dashed
line), where
\begin{equation}
C_{harm}=k_{B}4{\alpha}^{2}\frac{e^{2\alpha}}{\left( e^{2\alpha}-1\right)
^{2}}.\label{Charm}%
\end{equation}
The effects of the anharmonicity are strongest for values of ${\alpha}\leq0.5
$, where $\alpha=0.5$ corresponds to the characteristic vibrational
temperature $\Theta$ ($\Theta=4300$K for H$^{1}$Cl$^{35}$).

The graph shows an anomaly in the dependence of the vibrational specific heat
on the parameter $\alpha$ (temperature $T$). The specific heat has a maximum
for a value of ${\alpha}={\alpha}_{C}$ which corresponds to a temperature,
$T=T_{C}$. We shall call this temperature the critical temperature and the
corresponding parameter $\alpha_{C}$ the critical parameter. The anomaly of
the specific heat is again a result of the finite number of states in the
system. The specific heat increases with the temperature as more anharmonic
bosons are excited. The maximum is reached when the latter occupy the energy
state with $v=N_{0}=28$. (Note that the shape of the curve is similar to the
shape associated with the Schottky anomaly of the specific heat of a two-level
system \cite{BS:statmech}).

The maximal vibrational energy is given by the equation (\ref{Ev}) by
substituting $v$ with its maximal value $N_{0}$. Thus,
\begin{equation}
\varepsilon_{\max}=\hbar\omega_{0}\left(  \frac{N_{0}}{2}+\frac{1}{2}\right)
~~.\label{Emax}%
\end{equation}
while the minimum energy is,
\begin{equation}
\varepsilon_{0}=\frac{1}{2}\hbar\omega_{0} ~~.\label{Emin}%
\end{equation}
Thus,
\begin{equation}
\Delta\varepsilon=\varepsilon_{\max}-\varepsilon_{0}=\frac{1}{2}\hbar
\omega_{0}N_{0}%
\end{equation}
This gives $\Delta\varepsilon=14\hbar\omega_{0}=5.1877$eV for the molecule
H$^{1}$Cl$^{35}$. Comparing $\Delta\varepsilon$ with the dissociation energy
of the molecule $DE=4.4703$eV \cite{CRC}, we can conclude that at the
temperature $T=T_{C}$, $\Delta\varepsilon>DE$ and some of the molecules might
have started to dissociate while others may still be in stable molecular
states. Our model, in its present form, does not account for the effects of
the dissociation. In addition, this simple version of the model does not yet
include the contributions of the translational and rotational degrees of
freedom which at temperatures close to $T_{C}$ may be substantial. The
critical temperature $T_{C}$ can be considered as a temperature above which
the model is no longer valid in its current form and other effects take place,
{.g.} dissociation \cite{ADF:prep}.

\begin{figure}[t]
\input{epsf}  \centerline{\epsffile{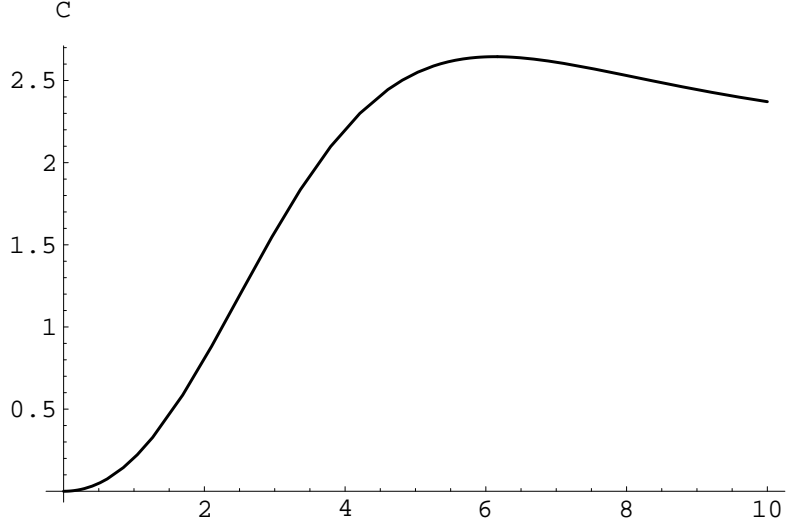}} \caption{Vibrational specific
heat $C_{N}/k_{B}$ as a function of ${\alpha}N_{0}$. }%
\label{fig:heatgen}%
\end{figure}

We have studied the behaviour of the specific heat with respect to the
combined parameter $\alpha N_{0}$. The graph of $C_{N}/k_{B}$ as a function of
${\alpha}N_{0}$ shows a similar anomaly (Fig. \ref{fig:heatgen}).

Solving numerically the equation $\frac{\partial{C_{N}}}{\partial({\alpha
}N_{0})}=0$ with respect to the combined parameter $\alpha N_{0}$, we have
found a root, $\alpha_{C}N_{0}=6.1332642 $. Thus, the critical value
$\alpha_{C}$ decreases as the number of fixed anharmonic bosons increases,
\begin{equation}
\alpha_{C}=\frac{6.1332642}{_{N_{0}}}\label{crit}%
\end{equation}
When $N_{0}~$ $\rightarrow\infty$, $\alpha_{C}\rightarrow0$ and the anomaly of
the specific heat disappears, which is in agreement with the harmonic limit of
the model.

For the molecule H$^{1}$C$^{35}$, $N_{0}=28$ which gives a value for
$\alpha_{C}=0.219$. Thus, the value of the critical temperature for this
molecule is $T_{C}=9815K$.

\subsection{Free Energy}

The free vibrational energy in terms of the partition function $Z_{N}$is given
by
\begin{equation}
F_{N}=-\frac{1}{\beta}\hbox{\rm ln} Z_{N}\label{FE}%
\end{equation}
Substituting $Z_{N}$ with equation (\ref{pf}) gives the free vibrational
energy in the algebraic model at high temperatures,
\begin{equation}
F_{N}=\frac{\hbar\omega_{0}}{2}\left[  \frac{1}{\alpha}\hbox{\rm
ln}
2+\frac{1}{2\alpha}\hbox{\rm ln} \left( \frac{\alpha}{\pi N_{0}}\right)
+\left(  N_{0}+1\right)  -\frac{1}{\alpha}\hbox{\rm
ln} \left(
\hbox{\rm erf} \,i\left(  \alpha N_{0}\right)  \right)  \right] \label{FN}%
\end{equation}
When $N_{0}~$ $\rightarrow\infty$, using expression (\ref{Zhl}) in equation
(\ref{FE}), we obtain the classical harmonic result for the free vibrational
energy at very high temperatures,
\begin{equation}
F_{\infty}\sim k_{B}T\hbox{\rm ln} 2.
\end{equation}

\section{Anharmonic Bosons}

\subsection{Mean Number of Anharmonic Bosons}

The mean vibrational energy in the anharmonic model can be written in terms of
mean number\ $\langle\nu_{N}\rangle$ of anharmonic quanta, each with energy
$\hbar\omega_{0} $,
\begin{equation}
U_{N}=\hbar\omega_{0}\left(  \langle\nu_{N}\rangle+\frac{1}{2}\right)
\label{Uv}%
\end{equation}
Substituting $U_{N}$ by equation (\ref{UN}), we obtain $\langle\nu_{N}\rangle$
in terms of the partition function $Z_{N},$
\begin{equation}
\langle\nu_{N}\rangle=\frac{N_{0}}{2}+\frac{1}{4\alpha}-\frac{N_{0}e^{-\alpha
}}{4\alpha Z_{N}}.\label{vN}%
\end{equation}
Using the expression (\ref{pf}) in equation (\ref{vN}), we obtain the
high-temperature value,
\begin{equation}
{\label{v}\langle\nu_{N}\rangle=\frac{N_{0}}{2}+\frac{1}{4\alpha}%
-\sqrt{\frac{N_{0}}{4{\pi}{\alpha}}}\frac{e^{\alpha N_{0}}}{\hbox{\rm
erf}
\,i\left(  \sqrt{\alpha N_{0}}.\right)  }}%
\end{equation}

The harmonic limit is obtained from equation (\ref{v}) when $N_{0}%
\rightarrow\infty$ and $Z_{N}$ is given by the expression (\ref{Zhl}),
\begin{equation}
\langle\nu_{\infty}\rangle\sim\frac{k_{B}T}{\hbar\omega_{0}}.
\end{equation}
\begin{figure}[t]
\input{epsf}  \centerline{\epsffile{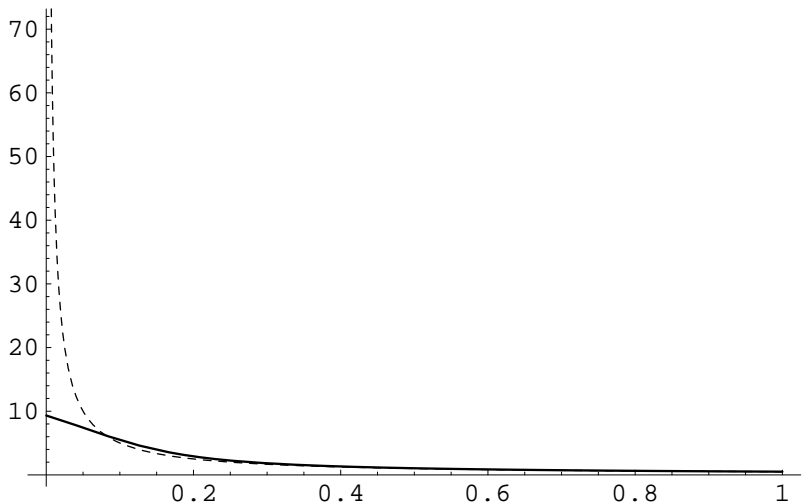}} \caption{Mean number of
anharmonic bosons $\nu_{56}$ as a function of $\alpha$. }%
\label{fig:bosons}%
\end{figure}

The graph of the function $\langle\nu_{56}\rangle$ for the molecule H$^{1}%
$C$^{35}$ is given on Figure \ref{fig:bosons}. The dashed line represents the
harmonic limit $\langle\nu_{\infty}\rangle$. The same reasons apply to the
finiteness of $\langle\nu_{56}\rangle$ as those discussed for the partition
function in Section 2.

\subsection{Maximal Temperature}

The maximal vibrational energy (\ref{Emax}) is obtained when $v=N_{0}$. The
maximal mean vibrational energy is given by equation (\ref{Uv}) when $\nu
_{N}=\langle\nu_{N}\rangle_{\max}$,
\begin{equation}
U_{\max}=\hbar\omega_{0}(\langle\nu_{N}\rangle_{\max}+\frac{1}{2}%
)~~.\label{Umax}%
\end{equation}
Comparing the equations (\ref{Emax}) and (\ref{Umax}) gives the maximal mean
number of anharmonic bosons,
\begin{equation}
\langle\nu_{N}\rangle_{\max}=\frac{N_{0}}{2}%
\end{equation}
(which according to the graph on Figure \ref{fig:bosons} is for a value
${\alpha}\rightarrow0$). Substituting $\langle\nu_{N}\rangle$ by equation
(\ref{v}) and simplifying gives,
\begin{equation}
2\sqrt{\frac{\alpha N_{0}}{\pi}}e^{\alpha N_{0}}={\hbox{\rm
erf} \,}i\left(
\sqrt{\alpha N_{0}}\right)
\end{equation}
The numerical solution of the above equation has a root $\alpha N_{0}%
\!\rightarrow\!0$. As $N_{0}$ is a fixed number, this solution leads to
$\alpha\rightarrow0$. In terms of the temperature, we obtain $T_{\max
}\rightarrow\infty$. The result shows that in practice the system does not
reach a maximal temperature, which shows the need of incorporating the
continuum into the partition function (\ref{def}).

\subsection{Quantum Anharmonic Bosons}

In \cite{angelova:JPA}, we have shown that the anharmonic bosons
$b,b^{\dagger}$ \ (\ref{qanh}) from Section 1 can be obtained as an
approximation of $q$-bosons
\cite{arik-coon:jmp,biedenharn:jphysa,macfarlane:jphysa}. The $q$-bosons are
defined by the following commutation relations:
\begin{equation}
\lbrack a,a^{\dagger}]=q^{\hat{n}}\ ,\quad\lbrack\hat{n},a]=-a\ ,\quad
\lbrack\hat{n},a^{\dagger}]=a^{\dagger}\label{qanh}%
\end{equation}
where the deformation parameter $q$ is in general a complex number
\cite{biedenharn:jphysa}. As shown in \cite{angelova:JPA}, the anharmonic
commutation relations (\ref{anh}) can be recovered for real values of the
deformation $q$ close to 1, $q<1$, and a linear expansion of $q$ in terms of a
parameter $p$, $p\equiv1/(1-q)$,
\begin{equation}
q^{\hat{n}}=1-\frac{\hat{n}}{p}~~.\label{qqanh}%
\end{equation}
If we now substitute the aproximation for $q^{\hat{n}}$ from equation
(\ref{qqanh}) in the commutation relations (\ref{qanh}) and identify the
parameter $p$ with $N/2$, $\hat{n}$ with $\hat{v}$ and the creation and
annihilation operators $a,a^{\dagger}$, with $b,b^{\dagger}$ respectively, we
recover the $SU(2)$ anharmonic relations (\ref{anh}).

The form (\ref{anh}) of the $SU(2)$ commutation relations can be considered as
a deformation of the usual (harmonic oscillator) commutation relations, with a
deformation parameter $p=N/2$. This gives a possible physical realization for
the quantum deformation obtained in \cite{angelova:JPA}: the quantum
deformation parameter $p$ is the fixed number $N_{0}$ of the anharmonic bosons
in the oscillator. Using the relation between the fixed number of anharmonic
bosons $N$ and the characteristics of the Morse potential (\ref{ND}) we arrive
to the conclusion that the quantum deformation is also determined by the
depth, the width and in general the shape of the Morse potential well. For the
molecule H$^{1}$C$^{35}$, $p=28$ which gives $q=27/28$.

Now, substituting $N_{0}=p$ in the expressions for the partition function
(\ref{pf}), mean energy (\ref{U}), specific heat (\ref{C}), free energy
(\ref{FN}) and mean number of anharmonic bosons (\ref{vN}), we obtain the
thermodynamic properties of diatomic molecules in terms of the deformation
parameter $p$. Equation (\ref{crit}) gives the relation between the quantum
deformation parameter and the critical parameter $\alpha_{C}$ (critical
temperature $T_{C}$). For large values of $p$, ($q{\rightarrow}1$), the
classic harmonic case is restored.

\section{Conclusion}

We have studied the vibrational thermodynamic properties of diatomic molecules
which at high temperature may strongly depend on its anharmonicity. We have
derived and studied the vibrational partition function and the related
thermodynamic functions, such as mean vibrational energy, specific heat and
mean number of anharmonic bosons, in terms of the parameters of the model. We
have also shown that it is possible to interpret these results in terms of a
quantum deformation, related to the shape of the Morse potential, and which is
associated with the fixed total number of anharmonic bosons, so that the
thermodynamic properties of diatomic molecules depend on the corresponding
quantum deformation parameter. We believe that this paper constitutes a first
step in the description of thermodynamical properties of diatomic molecules
which in principle can be simply generalized to polyatomic molecules. We are
currently studying the introduction of the continuum into the description, in
order to take into consideration the transition to dissociation
\cite{ADF:prep}.

\section{Acknowledgements}

Maia Angelova thanks Martin Levy, Alan Jones, Derek Gardiner and Jean-Pierre
Gazeau for useful discussions. This work was supported in part by Conacyt
project 32 397-E.

\end{document}